\documentclass[letterpaper,12pt]{report}
\usepackage{fullpage}
\usepackage{amsmath, amsthm, amssymb}
\usepackage{graphicx}
\usepackage{verbatim}
\usepackage{fancyhdr}
\usepackage[colorlinks=true,linkcolor=blue]{hyperref}
\usepackage[all]{xy}

\theoremstyle{plain}
\newtheorem*{thmHop}{The hopping immunity theorem}
\theoremstyle{definition}
\newtheorem*{example}{Example}

\graphicspath{{Images/}}


\title{\textbf{Analysis of Bitcoin Pooled Mining Reward Systems}}

\author{Meni Rosenfeld\\
}

\date{\today}

\begin{document}

\maketitle
\begin{abstract}
In this paper we describe the various scoring systems used to calculate rewards of participants in Bitcoin pooled mining, explain the problems each were designed to solve and analyze their respective advantages and disadvantages.
\end{abstract}
\pagenumbering{roman}
\tableofcontents
\chapter{Introduction}\label{chap:intro}
\pagenumbering{arabic}
\section{Bitcoin and mining}
Bitcoin is the world's first decentralized digital currency (\cite{Bitcoin}). It relies, among other things, on a network of computers that synchronize transactions with a process called \emph{mining}. Mining consists in repeatedly computing hashes of variants of a data structure called a \emph{block header}, until one is found whose numerical value is low enough. When this happens it allows releasing a valid \emph{block}, for which the miner is rewarded with bitcoins in an amount we will denote $B$.

Assuming sufficient quality of the hash function in the protocol and the pseudo-random number generator used in the construction of the block header, whether a given calculated hash leads to a valid block can be considered for all purposes a random event, independent of the validity of any other calculated hash. A quantity known as the \emph{difficulty} (which we will denote $D$) is adjusted periodically by the network and determines, as the name suggests, the difficulty of finding a valid block. The target value is chosen so that every computed hash will lead to a valid block with probability $\frac{1}{2^{32}D}$.\footnote{Strictly speaking it is actually $\frac{2^{16}-1}{2^{48}D}$. We will ignore this distinction throughout the paper.}

A miner with hashrate $h$ mining for a period of time $t$, will calculate a total of $ht$ hashes, and so
will find on average $\frac{ht}{2^{32}D}$ blocks. His expected payout is thus $\frac{htB}{2^{32}D}$.

\begin{example}Bob buys a dedicated mining computer which can perform a billion hash calculations per second, $h=1\,\textrm{Ghash/s}=10^9\,\textrm{hash/s}$. If he uses it to mine continuously for a day (86,400 seconds) when the difficulty is $D=1690906$ and the block reward is $B=50\,\textrm{BTC}$, he will find on average $\frac{ht}{2^{32}D}=\frac{10^9\,\textrm{hash/s}\cdot86400\,\textrm{s}}{2^{32}\cdot1690906} \approx 0.0119$ blocks, and receive payment of $0.0119B = 0.595\,\textrm{BTC}$ on average.\end{example}
\section{Variance of solo mining}
The process of calculating hashes individually, in order to find a valid block whose reward will be paid entirely to the person in ownership of the hashing computer, is known as \emph{solo mining} (to be contrasted with \emph{pooled mining}, introduced later).

Block finding when mining solo with a constant hashrate $h$ is a Poisson process with $\frac{h}{2^{32}D}$ as the rate parameter. We said that mining for time $t$ results in $\frac{ht}{2^{32}D}$ blocks on average. We can say further that number of blocks found follows the Poisson distribution with $\lambda=\frac{ht}{2^{32}D}$, so this quantity is also the variance\footnote{The \emph{variance} of a random quantity is a measure of its tendency to be different from its expected average. The standard deviation is the square root of the variance and is measured in the same units as the quantity in question.} of the number of blocks found. The variance of the payout is then $\lambda B^2=\frac{htB^2}{2^{32}D}$, and the relative standard deviation, as a fraction of the expected reward, is $\frac{\sqrt{\lambda B^2}}{\lambda B}=\frac{1}{\sqrt{\lambda}}=\sqrt{\frac{2^{32}D}{ht}}$.

\begin{example}Bob from the previous example has a variance of $0.0119B^2=29.75\,\textrm{BTC}^2$ in his payout. The standard deviation is $\sqrt{29.75\,\textrm{BTC}^2}\approx 5.454\,\textrm{BTC}$, which is 917\% of the expectation. In fact, the probability that Bob will receive any payment at all for his day of mining is $1-\exp(-\lambda)\approx 1.18\%$.\end{example}

We can see that the variance of solo mining is substantial. Even a participant with respectable hardware will wait about 3 months on average to receive any payout at all. And one should also note that the process is completely random and memoryless -- If the user has gone 3 months without finding a block, he isn't any closer than he was in the beginning and must wait on average 3 more months. Furthermore, the difficulty is expected to rise as Bitcoin gains popularity, which will only increase the variance.

This is problematic for several reasons:
\begin{itemize}
\item {Because the utility of money is concave, variance in the supply of money creates a direct decrease in a person's utility. It also makes it harder to make financial plans.}
\item{The lack of regular payments could make it technically more difficult to verify that all systems are working correctly.}
\item{It can be very taxing emotionally to have such a high-variance source of income.}

\end{itemize}

\section{Pooled mining} \label{sec:intro-pool}
A mining pool is a joint effort between several miners to work on finding blocks together, and split the rewards among the participants in proportion to their contribution. If the total hashrate of all miners is $H$, then the
pool will find on average $\frac{Ht}{2^{32}D}$ blocks over time $t$, so the total average reward is $\frac{HtB}{2^{32}D}$. A single miner with hashrate $h=qH$ (that is, $q$ is the fraction of the pool's total power contributed by this miner) should receive $q$ times the total reward, $q\frac{HtB}{2^{32}D}=\frac{htB}{2^{32}D}$, which is exactly his expected payout had he mined solo. However, his variance will be much lower -- the variance of the total reward is $\frac{HtB^2}{2^{32}D}$, so the individual's variance is $q^2\frac{HtB^2}{2^{32}D}=q\frac{htB^2}{2^{32}D}$, which is $q$ times the miner's solo variance. The potential benefit is therefore greater the smaller the miner and the larger the pool.

The pool will typically be maintained by a pool operator who may take a fee for his services. This will usually be a fixed percentage cut $f$ of the block reward. So, for every block found, the operator will receive a fee of $fB$ and the remaining $(1-f)B$ will be distributed among the miners. Therefore, the actual expected payout for a miner is $\frac{(1-f)htB}{2^{32}D}$.

To determine the amount of work done for the pool, users find and submit \emph{shares}, hashes of a block header which are low enough to have made a block if the difficulty was 1. Each hash has a probability of $\frac{1}{2^{32}}$ to be a share. Assuming correctness of the hash function used, it is impossible to find shares without doing the work required to find blocks, or to look for blocks without finding shares along the way. The number of shares found by a miner is therefore proportional, on average, to the amount of hashes the miner calculated in an attempt to find a block for the pool.

Every share has a probability of $p=\frac1D$ to be a valid block. Hence, if a miner spends solo mining the amount of effort it takes to find 1 share, his expected payout for it is $pB$. By spending this effort mining for the pool, his expected contribution to the total reward of the pool is $pB$. Hence, in a fair pool, the miner should receive on average $pB$ for every share submitted, or $(1-f)pB$ taking into account the operator's fee.

It turns out that deciding how to divide the rewards, so that each miner will be paid his fair share in proportion to his work, is not a trivial problem. A variety of methods have been devised for this purpose, some good, and some not so good. In the following chapters we will examine each such method in detail.

As noted above, a pool has the potential to improve the variance of a miner by a factor of $q$, the miner's part of the pool's hashrate. However, if the miner's hashrate is too small, this potential will not be fulfilled because of the variance in the number of shares he finds. There is a maximal possible improvement factor which depends on the details of the reward system used. Mining intermittently can also increase the variance, in a way which also depends on the reward system. However, these considerations have no effect on the more important quantity of the average payout, which in a fair pool is always $(1-f)pB$ for every share submitted.

\chapter{Simple reward systems}
\section{Proportional} \label{sec:prop}
The proportional system is perhaps the simplest pooled mining reward system, and the one that, intuitively, seems to best capture the principles of pooled mining set forth in \autoref{sec:intro-pool}.

In this system, payments are calculated based on a division to \emph{rounds}, where a round is the time between one block found by the pool to the next. At the end of every round, when a block is found and the pool receives a reward of $B$, the operator keeps a fee of $fB$, and $(1-f)B$ are distributed among the miners, in direct proportion to the number of shares they submitted during this round. If a miner submitted $n$ shares in this round, and the total number of shares submitted to the pool during this round is $N$, then his payout for this round will be $\frac{n}{N}(1-f)B$.

If the roster of miners, and their respective hashrates, are fixed, then the number of shares submitted by each miner in a round will be on average proportional to his hashrate. It follows that the expected payout for every share submitted is $(1-f)pB$. The variance in the payout for every share submitted is roughly $p^2B^2 \ln D$. These results are derived in \autoref{app:prop}. Note that for solo mining, the variance per share is $pB^2$, so this is an improvement by a factor of $\frac{D}{\ln D}$. This is only relevant for small miners, though; for miners that constitute a significant part of the pool, the payouts of shares will be correlated, and the total variance cannot be lower than $q$ times the solo variance. ($q=\frac{h}{H}$ is the miner's portion of the pool.)

Unfortunately, these results are only valid under the assumption of a fixed miner base. It turns out that a miner can, by strategically selecting when to mine for the pool and when to direct his hashrate elsewhere, receive a reward above and beyond his fair share for his contribution (which is $(1-f)pB$ per share on average), leading those who mine continuously to earn less than their due reward. This practice, known as \emph{pool-hopping}, was analyzed in \cite{Raulo} and is further analyzed in \autoref{app:hop}.

To understand intuitively why pool-hopping works, we should recall that the payout given for every share (disregarding fees for simplicity) is $\frac{B}{N}$. The longer the round, the less each share is worth. It is therefore to one's advantage if he could submit shares to this pool during its shorter rounds, and elsewhere in its longer rounds. Nobody can predict the future, but the past is not as mysterious; the number of shares \emph{already submitted} during this round, at the time of deciding on a course of action, directly affects the estimates of what the eventual length of the round will be.

For example, if already $2D$ shares have been submitted, then clearly by round end we will have $N\ge2D$, so the payout per share will be less than $pB/2$. It is definitely unprofitable to mine for the pool when its current round has dragged on for this long. And it is easy to see that, if mining throughout the round leads to a normal reward, and mining only late in the round leads to less than normal reward, then it must be the case that mining only early in the round leads to higher than normal reward. It turns out that the point where the expected payout is equal to the fair average payout, is when the number of shares already submitted is 43.5\% the difficulty. A pool-hopper will thus strive to mine only earlier than that, and ``hop'' elsewhere when this threshold is reached -- only to return when a new round starts.

In fact, we can point out what goes wrong with the proportional method's intuitive appeal. This intuition is perfect for projects of a more or less deterministic nature, whose time to completion can be estimated in advance with some certainty. Any work done on such a project advances us towards our goal, shortening the time remaining until completion. In such a project it only makes sense to reward each contributor in proportion to his work on a given part.

However, not so is the process of block finding. It is completely random. The number of shares in a round follows the geometric distribution (with success parameter $p$ and mean $D$), which is memoryless. There will be on average $D$ shares in a round. If some number $I$ of shares have already been submitted, the number of shares remaining will not be any less, it will follow the exact same distribution -- so on average $D$ \emph{more} shares will need to be found! Any newly submitted share will compete with $I$ previous shares and on average $D$ future shares, so to improve one's odds he may as well make sure $I$ is as low as possible. This is a probabilistic process, and must receive a probabilistic treatment and a probabilistic definition of fair division of work -- which is, that at all times, the expected payout per share is equal to the expected contribution.

The consequence of all this is that honest miners, who participate in the pool continuously hoping to earn their due reward, could theoretically in the worst case receive 43\% less. This is of course completely unacceptable. In practice the decrease will be much less, and depend on the number of pool-hoppers and the effectiveness of their techniques. However, the advent of tools such as bithopper (\cite{bithopper}) suggest that this is a significant problem.

Some pools choose to keep the proportional method but take various actions to defend against pool-hopping. These tend to have limited effectiveness and may end up harming honest users. The realization that the proportional system is inherently flawed has led to the development of score-based methods, some of which can be mathematically proven to be immune to pool-hopping and reward all miners fairly -- and these are the subject of \autoref{chap:score}.

\section{Pay-per-share (PPS)}
In the PPS system, the operator is not a passive middleman between the participants, coordinating the joint effort to reduce individual variance. Rather, he actively absorbs all of the variance each miner is facing. When a participant submits a share, he is immediately rewarded with $(1-f)pB$, corresponding to the expected value of this share's contribution, minus fees -- no matter how many blocks are eventually found. The operator gets to keep all the rewards for found blocks.

The payment per share is thus a deterministic value known in advance. This has several advantages for miners:
\begin{itemize}
\item {Zero variance in the reward per share. There is still some variance in the number of shares found by the miner in unit time, but this is mostly insignificant.}
\item{No waiting time until a block is found to obtain payment.}
\item{Easy to describe the exact payment that will be received.}
\item{Easy to verify that the promised reward is given, and that there are no losses due to dishonesty on the part of the operator or other parties.}
\item{No losses due to pool-hopping, which is ineffective against this method.}
\end{itemize}

However, this is the riskiest reward system for the pool operator -- to be able to offer zero variance to participants, he must take all the variance himself. He can make a nice profit on short rounds -- where he gains the entire block reward but only has to give payment for less than the average number of shares -- but can lose substantially on long rounds. His variance is the same as the solo variance of mining with the capacity of the entire pool (in absolute terms, the variance increases proportionally to the hashrate). To compensate for his risk, the operator will charge a higher fee than with other methods, and this is the disadvantage of PPS for miners.

If the operator doesn't correctly balance the pool's fee with his financial reserves, the pool has a good chance of eventually going bankrupt. As derived in \autoref{app:net}, to keep the bankruptcy probability below $\delta$, the operator should keep a reserve of at least
\[R=\frac{B\ln\tfrac1\delta}{2f}.\]

The required reserves are higher than most people anticipate. Operating such a pool is thus best left for those who know how to manage their risks responsibly, and miners who appreciate stability should shy away from improperly managed PPS pools which could shut down at any minute.

\chapter{Score-based methods} \label{chap:score}
\section{Slush's method} \label{sec:slush}
The first method intentionally designed to resist pool-hopping was implemented in slush's pool (\cite{slush}). It builds on the proportional method, but instead of basing participants' reward on a simple count of the number of shares they have submitted during the round, each share submitted credits the participant with a certain quantity of \emph{score}, and the block reward at round end is distributed among participants in proportion to their score.

The score given for each share depends on the amount of time that has elapsed since the round started. The more time has passed, the higher the score. Because in the proportional method shares submitted early are inherently worth more than late shares, this scoring counters the effect. The scoring function used is exponential, $s=\exp(T/C)$ where $s$ is the score given for a share submitted at time $T$, and $C$ is some constant.

An exponential function is adequate, because it is invariant under time shifts up to scale ($\exp((T+T')/C)=a\exp(T/C)$ where $a=\exp(T'/C)$). This means that some time after a round starts, a steady-state is reached where no matter when a share is submitted, its score relative to past and future shares is the same, and thus so is its expected reward.

The parameter $C$ controls how fast the score awarded for every share grows -- the lower it is, the faster the growth. Equivalently, it controls how fast the score of a given share decays relatively to new shares. If $C$ is low, the decay is fast -- this means that every share has a high chance to decay completely before a block is found, thus receiving no payment, while if a block is quickly found, the payment given will be very high because it will not be split between many other shares. The result is that a low value of $C$ significantly increases the variance of the received payments, while reducing the vulnerability to pool-hopping.

Despite the historical importance of this method as a pioneer in combating pool-hopping, it has several shortcomings:
\begin{itemize}
\item{A steady state is only reached some time after a round starts. At the beginning of a round, there aren't many previous shares with which a potential reward is to be shared, and thus it is more profitable to mine early in the round. The magnitude of the effect is not as large as in proportional pools, but the method is still not completely hopping-proof.}
\item{Because the score is determined by the amount of time elapsed, rather than the number of shares submitted, it is subject to hopping based on fluctuations in the pool's hashrate (discussed in \autoref{app:fluctuate}).}
\item{There is no consideration of the current difficulty in the score calculations, so it is hoppable based on expected difficulty adjustments.}
\end{itemize}
The first problem in this list is the most fundamental. This is an instance of a much more general phenomenon. The only way a reward system which divides a fixed reward per round among the participants in the round can be hopping-proof, is if it only rewards the share that solved the round (and is thus equivalent to solo mining, which is undesirable because of the high variance) -- this is the hopping immunity theorem, proven in \autoref{app:immune}. Thus, a hopping-proof method should defy either the ``fixed reward'' part, or the ``among the participants in the round'' part. In the rest of this chapter we will discuss two hopping-proof methods which do each of these, respectively.

\section{Geometric method}
The geometric method is a hopping-proof reward system inspired by slush's method. It, too, involves exponentially decaying score, but addresses the weaknesses of slush's method and puts the system on a solid mathematical framework.

In this method, there are two kinds of fee, a fixed fee and a variable fee. The fixed fee is a constant amount taken from the reward of every block. The variable fee takes the form of a score granted automatically to the pool operator at the start of every round, and decays just like the score of participants. Therefore, the shorter the round, the higher the variable fee. This is designed to create a steady-state where the score granted for every new share, relatively to already existing score and the score of future shares, is always the same, thus there is no advantage to mining early or late in the round.

The method goes as follows:
\begin{enumerate}
\item{Choose parameters $f$ and $c$.}
\item{At the start of every round, set $s=1$. For every worker $k$, let $S_k$ be the worker's score for this round, and set $S_k=0$.}
\item{Set $r=1-p+\frac{p}{c}$, where $p=\frac1D$. If the difficulty changes during the round, $r$ needs to be updated.}
\item{When worker $k$ submits a share, set $S_k=S_k+spB$ (where $B$ is the block reward at the time it was submitted), and then $s=sr$.}
\item{If the share is a valid block, end the round. For every worker $k$ pay $\frac{(1-f)(r-1)S_k}{sp}$.}
\end{enumerate}

$f$ is the fixed fee. If the block reward is a constant value $B$, then out of every block $fB$ will be kept by the operator and $(1-f)B$ will be distributed according to score.

$c$ is the average variable fee. Out of the $(1-f)B$ distributed per block, on average $c(1-f)B$ goes back to the operator as a variable fee. This means that out of every block a total of $(c+f-cf)B$ on average is paid to the operator and $(1-c)(1-f)B$ is paid to the participants. The total average fee is therefore $c+f-cf$.

$s$ is a counter keeping track of the score to be awarded for the next share to be submitted, expressed as a multiple of $pB$, the fair average. As the round progresses $s$ will grow exponentially, causing older shares to decay in their real value.

$r$ is the decay rate. The score for each share is $r$ times that of the previous one. If the difficulty is fixed, $r$ is constant so the score for the $I$th share submitted in the round is simply $r^{I-1}pB$. The slower the decay, the higher the operator's score needs to be to maintain a steady state, thus the higher the average variable fee is.

If $p$ and $B$ are fixed, then letting $N$ be the total number of shares submitted in the round, we have $s=r^{N}$ and a worker's payout is
\[\frac{(1-f)(r-1)S_k}{sp}=\frac{S_k}{r^{N}pB/(r-1)}(1-f)B=\frac{S_k}{\sum_{i=-\infty}^{N}r^{i-1}pB}(1-f)B\]
This is equivalent to distributing the $(1-f)B$ reward in proportion to score, where the operator gets a score of $\sum_{i=-\infty}^{0}r^{i-1}pB=\frac{pB}{r-1}$ and the total score of all participants is $\sum_{i=1}^{N}r^{i-1}pB$.

What this means is that the operator effectively gets infinitely many shares at the start of every round. Whenever a miner has the opportunity to submit a share, he sees behind him an infinite series of shares, with scores starting with the score of the share he will submit and decaying at a given rate. The statistical properties of his reward for this share are thus always the same.

The expected payout per share is $(1-f)(1-c)pB$. The expected fee is $(c+f-cf)B$ per block. The variance of a participant's payout, per share, is approximately $\frac{(pB)^2}{2c+p}$. For solo mining this would have been $pB^2$, so this is a decrease of roughly $1+\frac{2c}{p}$ times. The variance of the operator's fee, per block, is approximately $cB^2$. These results are derived in \autoref{app:geom}.

If the value $c+f-cf$, which is the average fee rate, is kept fixed while $c$ increases and $f$ decreases, the variance for the operator is increased and for participants it is decreased. As long as $f$ is positive, the operator will never lose money on a round. If the operator chooses to absorb more variance on behalf of the participants, he can let $f$ be negative and push $c$ further, effectively adding his own money to the block reward. This means he will lose up to $(-f)B$ per block on especially long rounds. He can even choose $f = (-c)/(1-c)$ and have an expected fee of 0.

In the limit $c\to0$, only the winning share is rewarded and the participants are effectively mining solo. If $c+f-cf$ is kept fixed and $c\to1$ (so $f\to-\infty$), the method reduces to pay-per-share (note that the approximate variances given above assume $c$ and $f$ are small and so don't apply for this case). $c$ shouldn't be outside the range $(0, 1)$.

Since $s$ grows exponentially, it will eventually overflow if the method is implemented naively. One of the following techniques should be used to keep the values of the variables small:
\begin{itemize}
\item{Periodically rescaling the scores. The quantity that really matters is the ratio between workers' scores and the current value of $s$. Nothing is changed if they are all divided by the same quantity. So, once in a while we should set $S_k=S_k/s$ for all workers, and then $s=1$.}
\item{Using a logarithmic scale. Instead of storing the actual values of $s$ and $S$, keep their logarithms, $ls$ and $lS$, and adapt all calculations to work with the new variables without overflowing. Implementing the method in a logarithmic scale goes as follows:
    \begin{enumerate}
    \item{Choose $f$ and $c$ as usual.}
    \item{At the start of every round, set $ls=0$. For every worker $k$, set $lS_k=-\infty$ (or some other very low value such as $-1000000$).}
    \item{Set $r=1-p+\frac{p}{c}$, and $lr=\ln(r)$. If the difficulty changes during the round, $r$ and $lr$ need to be updated.}
    \item{When worker $k$ submits a share, set $lS_k=ls+\ln(\exp(lS_k-ls)+pB)$, and then $ls=ls+lr$.}
    \item{When the round ends, for every worker $k$ pay $\frac{(1-f)(r-1)\exp(lS_k-ls)}{p}$.}
\end{enumerate}
    }
\end{itemize}

At any point, a worker's score indicates the expected reward for the shares he has already submitted. Specifically, the expected pending reward is $(1-f)(1-c)S_k/s$ (or $(1-f)(1-c)\exp(lS_k-ls)$ in logarithmic scale). This value can be displayed to participants to keep track of their earnings.

\section{Pay-per-last-N-shares (PPLNS)}\label{sec:pplns}
PPLNS is a family of methods that do away with the concept of ``rounds'' on which traditional pools are based. Instead of distributing block rewards among participants in the current round, that is, those who submitted shares after the previous block was found, they are distributed among those who submitted shares recently, regardless of whether any blocks were found in the examined period. By so doing, it removes the concept of profiting by mining ``early in the round'', and some variants of PPLNS methods are hopping-proof.

The simplest variant is to choose a fixed number $N$ and pay out a reward of $\frac{(1-f)B}{N}$ for each one of the last $N$ shares submitted. As long as $D$ and $B$ are constant, the expected payout per share is always $(1-f)pB$ -- the payout for a share is $\frac{(1-f)BL}{N}$, where $L$ is the number of blocks found within the next $N$ shares. Each share has a probability of $p$ to be a block, independently of the past, so $L$ follows the Poisson distribution with mean $\lambda=pN$, making the expected reward $\frac{(1-f)BpN}{N}=(1-f)pB$. The variance of $L$ is also $pN$, making the reward variance $\frac{(1-f)^2pB^2}{N}\approx\frac{pB^2}{N}$, which is $N$ times better than solo mining. The method has no effect on pool-based variance (the variance caused by the pool being too small for the miner), though.

For example, let $N=D$ and $f=0$. Then every share will be paid $pB$ (the fair average reward) for every block found in the next $N$ shares. The number of payments follows the Poisson distribution with $\lambda=1$, illustrated in \autoref{tab:PPLNS}. The average number of payments is 1.

\begin{table}
\caption{Distribution of number of payouts in PPLNS, for $N=D$.}
\begin{tabular}{c|r}
Number of&\\
payments&Probability\\
\hline
0& 36.79\%\\
1& 36.79\%\\
2& 18.39\%\\
3& 6.13\%\\
4& 1.53\%\\
5& 0.31\%\\
6& 0.05\%\\
\vdots& $<0.01\%$\\
\end{tabular}
\label{tab:PPLNS}
\end{table}

The timing of the payments is distributed uniformly over the period of the next $N$ shares, so each reward is received on average after $\frac{N}{2}$ shares, which in multiples of the difficulty is $\frac{pN}{2}$. The maturity time is therefore $\frac{pN}{2}$.

Increasing $N$ reduces the variance but increases the maturity time. The invariant of the tradeoff is that their product is always $\frac12(pB)^2$.

If we drop the assumption that $D$ and $B$ are fixed, this simple variant is not hopping-proof. A participant's contribution is determined by the current difficulty, while his reward is influenced by the future difficulty. Pool-hoppers can use knowledge of imminent difficulty adjustments to their advantage -- joining when the difficulty is set to decrease, and leaving when the difficulty is set to increase. Another simple variant, where the number of shares is chosen as a given multiple of the difficulty at the time a block is found, suffers from similar problems. To be completely hopping-proof, the following method, called ``unit-PPLNS'', should be used:

\begin{enumerate}
\item{Choose parameters $f$ and $X$. $f$ is the fee, $X$ is the window size measured in multiples of the average time to find a block.}
\item{When a share is submitted, store two attributes for it -- units, equal to the value of $p$ at the time it was submitted, and amplifier, equal to the value of $B$ at the time it was submitted.}
\item{When a block is found, let $u_1$ be the units of the winning share that solved the block. For every share except the winning one (the winning share does not receive a reward from its own block), let $u_0$ be its units, $a$ its amplifier, and $U$ the total units of all shares submitted after it (including the units of the winning share). Pay the participant who submitted this share
    \[\frac{(1-f)au_0\max\left(\min\left(\frac{X-U+u_1}{u_1},1\right),0\right)}{X}.\]}
\end{enumerate}

In practice, one way to implement this is by keeping track of $U^T$, the total units of all shares ever submitted, and storing with each share an indexed field $U^T_0$, the value of $U^T$ at the time it was submitted. When a block is found and the current value of $U^T$ is $U^T_1$, for every share we have $U=U^T_1-U^T_0$, so the only shares for which any payment is due are those for which $U^T_0>U^T_1-X-u_1$. Care should be taken to make sure that the data structures used for $U^T$ can handle the precision required.

The overall properties of unit-PPLNS are similar to those of simple PPLNS. The total payment for a share is $\frac{(1-f)pBL}{X}$, where $p$ and $B$ are the inverse difficulty and the block reward at the time it was submitted, and $L$ is the number of blocks found within the next shares with a total of $X$ units\footnote{Fractional payments are possible, and these are handled correctly and give rise to an accurate expected reward. However, in this analysis we use a continuous approximation and these are ignored.}. Every share has probability to be a block equal to its units, and thus $L$ follows the Poisson distribution with mean $\lambda=X$, independently of the past or any future difficulty adjustments. The expected reward per share is thus $(1-f)pB$, and the variance is roughly $\frac{(pB)^2}{X}$ (an improvement of $\frac{X}{p}$ times over solo). The maturity time is $\frac{X}{2}$, and the product of variance and maturity time is $\frac12(pB)^2$.

The expected pending reward for a share with amplifier $a$, units $u$ and a total of $U$ units submitted after it, is $\frac{(1-f)ua\max(X-U,0)}{X}$. The total of this over all the shares of a worker can be displayed as his pending reward.

Because there is no clean separation to rounds in PPLNS, an operator who wishes to change the parameters used does not have an opportunity to do so between rounds. Therefore, he must modify the units and/or amplifier of each share to make sure that its expected pending reward after the change is equal to the one before. If he does this by modifying the units, he must do so starting from the last share and moving backwards, since changing the units of one share changes the value of $U$ for all earlier shares. For changing $f$, it is simplest to scale the amplifier of every share by a factor of $\frac{1-f_1}{1-f_2}$. For changing $X$, it is simplest to scale the units of every share by $\frac{X_2}{X_1}$ and its amplifier by $\frac{X_1}{X_2}$; this way, the values of both $ua$ and $\max(X-U,0)$ for every share remain unchanged. If the operator chooses another strategy for altering share values, he may run into a situation (especially when $X$ is shortened) where for some share $U>X$ and hence no change in its values will maintain the pending reward. In this case he must instantly pay the expected pending reward. This represents paying back the statistical loan he took when he started the pool with a large value of $X$.

Some additional variants and generalizations of PPLNS are discussed in \autoref{sec:pplnsvar}.

\section{Score-based methods myths}
Reward systems such as the geometric method and unit-PPLNS are fair, stable, transparent, and hopping-proof. It should be clear that they are vastly superior to the proportional method which can be gamed to the detriment of honest miners, causing a substantial decrease in their long-term rewards. For reasons on which the author can only speculate, proportional pools are still rampant, and a lot of confusion and myths detract from the score-based methods. Here we will investigate some of them.
\begin{itemize}
\item{\textbf{``Pool-hopping does not work.''} The results in \autoref{app:hop} clearly indicate that pool-hopping in proportional pools indeed works and can be very profitable. Various thought experiments can help visualize why this must be so, simulations have verified the analytical derivations, and there is overwhelming empirical evidence by miners who profitably pool-hop.}
\item{\textbf{``Pool-hopping benefits hoppers without harming continuous miners.''} The total reward in a pool with a given average hashrate is constant. Any additional reward obtained by hoppers must be at the expense of continuous miners. \autoref{app:hop} quantifies the potential losses.}
\item{\textbf{``Pool-hopping is the efficient way to mine, everyone should do it.''} If everyone pool-hops, all proportional pool will eventually come to a standstill as they reach a number of shares in the current round equal to 43.5\% of the difficulty, at which point no-one will mine there anymore. Unlike other optimizations which increase the hashrate and thus the contribution to the pool and to the Bitcoin network's security, pool-hopping attempts to grab more rewards without giving anything back.}
\item{\textbf{``Score-based methods attempt to detect pool-hoppers and penalize them.''} Score-based methods do not encode in any way whether one is hopping or not. The scoring method is applied uniformly to all participants, and guarantee that their reward will be on average exactly proportional to the number of shares they submit over a period of time, whether they mine continuously, intermittently, or try to pool-hop.}
\item{\textbf{``The value of my score will decay if I quit a score-based pool, so it forces me to stay.''} Score and payouts are additive. The score you were rewarded for shares submitted in the past will decay the same way whether you stay or quit; the score you will receive for shares you submit in the future does not rely on the past. Your total payout is an independent sum of the two factors.}
\item{\textbf{``Score-based methods are disadvantageous for intermittent miners.''} The expected payout per share in a hopping-proof method is always the fair solo average no matter when it was submitted, so your expected total reward over a period of time is equal exactly to the total value of the shares you submitted, no matter what your mining pattern is. Intermittent mining does, however, increase the variance of the rewards, which is felt more strongly in high-variance systems like the geometric method. This will have very little effect in the long run, and even less so in PPLNS in which the variance is lower.}
\item{\textbf{``Mechanisms to detect and punish hoppers, employed by some proportional pools, work.''} Such methods are usually very poorly thought out. Generally they will have little effect on hoppers but are capable of harming honest miners.}
\item{\textbf{``Hiding information about the pool's found blocks is good.''} Some proportional pools attempt to deprive pool-hoppers of the data they need to successfully hop, delaying statistics on blocks found by the pool. This has some degree of effectiveness, but intelligent hoppers are able to hop to some extent even without access to this data. Also, pools should be transparent and hiding information is not a good practice. Since hopping-proof methods are available, there is no justification to resort to these techniques.}
\end{itemize}

\chapter{Attempts for risk-free pay-per-share}
\section{Maximum pay-per-share (MPPS)}
The attractiveness of the PPS method for miners, contrasted with the high risk it poses to pool operators, has led to attempts to offer rewards that mimic PPS but only pay out on a ``best effort'' basis according to the luck of the pool.

The first such method is MPPS. In this method, two balances are kept for each participant -- a PPS balance and a proportional balance. Whenever the participant submits a share, his PPS balance is increased as if this was a PPS pool. Whenever the pool finds a block, the proportional balance of each participant is increased as if this was a proportional pool. The total amount paid for each participant is the minimum between his PPS balance and his proportional balance.

The idea behind this system is that, since the total proportional balance of all participants is equal to the total reward obtained by the pool, and the total amount payed out is less than this quantity, the pool cannot lose money. And, since a participant's PPS balance is the accumulation of the fair average reward for all work done, he cannot use pool-hopping to have an expected reward higher than the fair average. Lucky (short) rounds do not pay out a high amount immediately, but rather build up the participant's buffer to help pay during unlucky rounds.

However, there are several blatant flaws with the method. First, the total amount paid is the minimum of two different quantities, each of which has an expected value equal to the fair average. It follows that the expected reward in this method is less than the fair average. In other words, in the PPS method, the normal reward is received no matter the luck of the pool; in the proportional method, the reward is higher than normal when the pool is lucky and lower than normal when the pool is unlucky; in MPPS, the reward is normal when the pool is lucky and lower than normal when the pool is unlucky.

We can quantify this effect -- in \autoref{app:MPPS} we show that if Alice joins an MPPS pool and mines for the amount of time it takes the pool to find on average $n$ blocks, then the average portion of the reward lost due to this method is $\frac{1}{\sqrt{2\pi n}}$. The losses become less significant relatively (though larger in absolute terms) the more time Alice spends in the pool. For example, if $n=10$, the average loss is $\frac{1}{\sqrt{20\pi}} \approx 12.6\%$.

Also, the method is not hopping-proof, as the expected reward per share depends on the time it is submitted. By submitting shares only early in a round, a pool-hopper can make sure his proportional balance is high, thus basing his rewards solely on his PPS balance. This way he effectively gets 0-fee PPS payments, which is better than alternatives (PPS with high fee, or low-fee pools with variance and maturity time). For honest miners, hopping causes their proportional balance to be less than expected, thus reducing their reward further.

\section{Shared maximum pay-per-share (SMPPS)}
The SMPPS method attempts to solve the problems of MPPS by replacing the per-user buffer with a global pool buffer shared by all users. Lucky rounds build up the pool buffer, which helps hand out PPS payment during unlucky rounds. When the buffer runs out, full PPS payments are delayed until more funds are collected. The exact way this is done depends on the specific SMPPS variant.

In the original SMPPS, the pool keeps track of each participant's due reward. Whenever he submits a share, his due reward is increased as in a PPS pool; whenever payment is handed out, it is deducted from the due reward. The pool also keeps track of its buffer $R$, defined as the difference between the total block reward ever earned by the pool and the total PPS worth of the work ever done by participants. Having $R>0$ means that the pool has earned enough to pay the participants for all work done, and has $R$ bitcoins to spare. When shares are submitted they are immediately paid in full out of this buffer. Having $R<0$ means there are participants to whom payment is due, but the pool does not have the funds to pay; the total due to all participants is $(-R)$. In this case, whenever a block is found, its reward is split among all participants in proportion to their due payment; that is, a participant with a due reward of $w$ is paid $\frac{wB}{-R}$.

The first problem with this method is that when $R<0$, the maturity time (the average time it takes to receive rewards) is high, and gets higher the lower $R$ is. In \autoref{app:MPPS} we show (in a slightly simplified model) that the maturity time is $\frac{-R}{B}$ (the pool's debt expressed as a multiple of the block reward). For example, if $R=-500\,\textrm{BTC}$ and $B=50\,\textrm{BTC}$, then the average time it takes to receive the reward is the time it takes the pool to find 10 blocks.

The second problem is that the buffer is guaranteed to reach arbitrarily low levels eventually. That is, if the pool continues operating indefinitely, there is a probability of 100\% that at some point its buffer will be $-1\,000\,\textrm{BTC}$; at some point it will be $-10\,000\,\textrm{BTC}$, and so on. This is because the pool's buffer follows a stochastic process similar to the one discussed in \autoref{app:net}, but with no average upward trend. In the best case this is equivalent to a one-dimensional Brownian motion, which is recurrent, and thus has the property of reaching any level with probability 1. When the debt becomes large enough, the maturity time will be so high that participants will barely see any payment given to them in practice. They will start leaving the pool, amplifying the problem since a decrease in the pool's hashrate will increase the time it takes to find a block, and hence the maturity time when measured in units of time. The pool will thus enter a downward spiral ultimately leading to its collapse. When this happens, everyone who has payment due will never receive it.

The third problem is that the collapse will actually happen sooner than would have been expected with the pure Brownian model. Inefficiencies such as stale shares (if paid), invalid blocks, sabotage and so on mean that the actual expected gain per share for the pool will be less than the payment due to the miner for it. This will cause a small but steady negative drift in the buffer.

Finally, the method is not hopping-proof. While the expected payout per share is, in theory, fixed, the maturity time is not. Mining for the pool when its buffer is positive is attractive since it is essentially 0-fee PPS. When the buffer is negative it is not attractive, because of the high maturity time. Pool-hoppers will mine when the buffer is positive and leave when it is negative. Honest miners willing to put up with the high maturity time in bad times in return for the steady payments in good times, may find that the hoppers are the primary beneficiary of the good times, while they are left suffering from more than their fair share of the bad times.

\section{Equalized SMPPS (ESMPPS)}
ESMPPS is a variant of SMPPS based on the idea that some minimal reward per share should be given as soon as possible, while reaching the full payment is not as important.

In this method, the pool keeps a record of all submitted shares and tracks how much payment was given for each. As long as the pool has enough funds to pay each share in full it does. If not, then when a block is found, its reward is distributed among shares so as to maximize the minimum percentage paid over all shares. In other words, the shares which so far were paid the least are prioritized.

Because of this, even if the total due payment is very large, new shares can quickly reach the payment level of older shares. But once a share has reached a relatively high payment level, luck is required to gain any additional payment at all, so the approach to 100\% payment is very slow. In fact, the developers of the method do not consider striving for 100\% to be a realistic goal, and mostly consider the time it takes to approach the 97\% level.

It can be shown\footnote{The buffer, evaluated against the efficiency level, follows Brownian motion. With probability 1, it will eventually be positive meaning full efficiency payment; but the expected time to reach any lower level is infinite, and this is a lower bound for the time of the payment between the two levels.} that the maturity time of this method is infinite. More than anything this result indicates that our notion of maturity time is too crude to describe the behavior of this method, but it still follows that for a wide range of choices it is inferior in this regard to SMPPS with the same buffer.

The advantage of ESMPPS is that it handles losses gracefully. Inefficiencies will reduce the payment level to which shares approach, but the buffer evaluated against this target level does not have a negative drift. This outcome comes about organically and doesn't require to know the loss level in advance. However, even if the buffer follows plain Brownian motion, it is still guaranteed to eventually reach arbitrarily low levels. And, like SMPPS, this method is not hopping-proof.

\chapter{Advanced methods} \label{chap:advanced}
\section{Double geometric method}
Every hopping-proof reward system requires maintaining a steady-state where every miner always sees the same relative history no matter when he submits a share. The geometric method does this by taking a variable fee -- if there are not enough shares in the current round, the operator emulates such shares with his fee. Rewards are cleanly separated into rounds -- every share is rewarded only for the next block found and not for any later ones. Reducing variance this way requires the operator to absorb it, increasing his risk -- but in so doing, not only share-based variance is reduced (variance caused by discontinuities in submitting shares), but also the pool-based variance (variance caused by the pool being too small).

PPLNS maintains a steady state by ignoring rounds completely -- the reward given to a share from a block is completely independent of any blocks found in between, and the recent history is always a given number of shares submitted by participants. Lengthening the window allows decreasing the share-based variance arbitrarily with no risk for the operator (at the cost of increased maturity time). However, the method offers no mechanism to reduce pool-based variance.

The double geometric method is a hopping-proof reward system that falls between these two extremes. Round boundaries are crossed, but not ignored -- every block found reduces the reward to be given for future blocks, but does not erase it completely. A parameter controls the degree of reduction, thus setting the exact location on the spectrum between PPLNS and geometric.

The parameters used by the method to adjust the balance between average fee, operator variance, share-based and pool-based participant variance, and maturity time are:
\begin{itemize}
\item{$f$ is the fixed fee. In conjunction with the other parameters it determines the average fee collected by the operator.}
\item{$c$ is the average variable fee. The average total fee will be $(c+f-cf)B$ per block. Increasing $c$ reduces participants' variance but increases the operator's variance.}
\item{$o$ is the cross-round leakage. Increasing $o$ reduces participants' share-based variance but increases maturity time. When $o=0$ this becomes the geometric method. When $o=1$ this becomes a variant of PPLNS, with exponential decay instead of a step function. In the case $o=1$, $c$ must be 0 and $r$ (defined below) can be chosen freely (to control the decay rate) instead of being given by a formula.}
\end{itemize}
With the method description used here, the parameters can be changed at any point without affecting the expected payout for already submitted shares.

The method goes as follows.
\begin{enumerate}
\item{When the pool first starts running, initialize $s=1$. For every worker $k$, let $S_k$ be the worker's score, and set $S_k=0$.}
\item{Set $r = 1 + \frac{p(1-c)(1-o)}{c}$. If at any point the difficulty or the parameters change, $r$ should be recalculated.}
\item{When worker $k$ submits a share, set $S_k = S_k + (1-f)(1-c)spB$ (where $B$ is the block reward at the time it was submitted), and then $s = sr$.}
\item{If the share is a valid block, then after doing \#3, also do the following for each worker $k$: Give him a payout\footnote{Or equivalently, a payout of $\frac{(r-1)S}{(1-c)ps}$. This is useful if $o=1$.} of $\frac{(1-o)S_k}{cs}$, and then set $S_k=S_ko$.}
\end{enumerate}
The intuition is this: Instead of keeping the score unchanged when a block is found (as in PPLNS) or setting all scores to 0 and effectively transferring them to the operator (as in the geometric method), a part of the score is transferred to the operator. When rounds are long, participants get to keep most of their score between rounds and this is similar to PPLNS. However, if several blocks are found in rapid succession, the operator will collect a large portion of the score and thus be the primary beneficiary of the good fortune. The fees collected this way allow letting f be negative, sweetening the rewards of long rounds. Overall, this decreases the dependence of participants' rewards on the pool's luck, thus reducing the variance caused by it.

As with the geometric method, implementation will require either periodically rescaling the scores or using a logarithmic scale. Implementing the method with a logarithmic scale is as follows:
\begin{enumerate}
\item{When the pool first starts running, initialize $ls=0$. For every worker $k$, set $lS_k=-\infty$ (or some other very low value such as $-1000000$).}
\item{Set $r = 1 + \frac{p(1-c)(1-o)}{c}$ and $lr=\ln(r)$. If at any point the difficulty or the parameters change, $r$ and $lr$ should be recalculated.}
\item{When worker $k$ submits a share, set $lS_k = ls + \ln(\exp(lS_k-ls) + (1-f)(1-c)pB)$, and then $ls = ls+lr$.}
\item{If the share is a valid block, then after doing \#3, also do the following for each worker $k$: Give him a payout of $\frac{(1-o)\exp(lS_k-ls)}{c}$, and then set $lS_k=lS_k+\ln(o)$.}
\end{enumerate}
At any point, the expected pending reward of worker $k$ is $S_k/s$ (or $\exp(lS_k-ls)$ in logarithmic scale).

\section{General unit-based framework}\label{sec:framework}
Revisiting the PPLNS and geometric methods, we can identify two ways by which they differ:
\begin{itemize}
\item{Round separation: In the geometric method, rounds are cleanly separated; in PPLNS, round boundaries are completely ignored.}
\item{Decay function: The geometric method uses an exponential function; shares decay exponentially as more shares are submitted. PPLNS uses a step function; a share is either counted fully if it is in the window, or disregarded completely if it is outside it.}
\end{itemize}
The two attributes are independent. We could just as well have a hopping-proof method with round separation and step decay, or with exponential decay and no rounds (the latter is what DGM becomes for $o=1$). We could also have partial round separation (as in DGM), or even use a completely different decay function. In this chapter we develop a general framework for dealing with this family of methods. The general method reduces almost exactly to the geometric method, PPLNS and DGM as special cases. However, since the framework is continuous while the reality of mining is discrete, applying it directly results in extremely negligible dependence of profitability on difficulty changes.

Every method in the family is characterized by two parameters\footnote{In this discussion we ignore the fixed fee $f$, which can be used to adjust the average fee.}, a round separation parameter $O$ and a decay function $r(x)$. $O$ is a nonnegative number and $r$ is a function satisfying the following properties:
\begin{enumerate}
\item{$r(x)\ge0$ for every $x$.\footnote{The framework works without this condition, but allowing negative values would require a two-way payment channel between the operator and participants.}}
\item{$r(x)=0$ for every $x<0$.}
\item{$\int_0^{\infty}r(x)\ dx=1$.}
\end{enumerate}
Typically, $r$ is a monotonic decreasing function. For example, for exponential decay we will use $r(x)=[x\ge0]\cdot\alpha\exp(-\alpha x)$;\footnote{$[P]$ is the Iverson bracket, $[P]=\left\{\begin{array}{cc}1&\textrm{if $P$ is true;}\\0&\textrm{otherwise.}\end{array}\right.$} for step decay we will use $r(x)=[0\le x\le X]\cdot(1/X)$; and for linear decay we will use $r(x)=[0\le x\le X]\cdot2(X-x)/X^2$.

The idea is that for every share, every block found gives a reward of $r(x)$, where $xD$ is the number of shares found between the share and block in question. To this end, we construct a ``timeline'' of shares, where the shares submitted are lined up in order and each occupies a width of $p$ ``units'', where $p$ was the value at the time the share was submitted. In addition each share is characterized by its amplification factor, equal to the value of $B$ at the time it was submitted.

When a share is found which is also a valid block, it is (tentatively) added to the timeline normally, and then rewards are given to all previous shares. Rewards are computed based on the decay function and the distance on the timeline between the block and each share. Since every share, including the block, has positive width rather than occupying a single point on the timeline, we treat it as a superposition of many infinitesimal subshares. Finding the exact reward requires integration over both the block and the paid share. Denoting by $A$ the amplification of the paid share, by $p_1$ the paid share's units, by $p_2$ the found block's units, and by $x$ the distance between the end of the paid share and the beginning of the block share (the block share receives payment itself, with $x=-p_2$), the payment is
\[\frac{A}{p_2}\int_0^{p_1}\int_0^{p_2}r(x+u+v)\ du\ dv.\]
After every block found, a void of width $O$ units is added to the timeline. Any new shares and blocks will be added after this void. This causes decay of current shares, and is equivalent to virtual shares credited to the operator. If $O=0$, blocks do not cause any decay, as in PPLNS; if $O=\infty$, blocks cause existing shares to reset completely, as in the geometric method.

Some care should be taken with regards to rewards for the winning share from future blocks. Since the actual block is treated as if found in a random point within the width of the share, it is not correct to put the avoid entirely after it -- rather, the continuum of possibilities should be superposed. This is done by splitting the share in two -- a share of $p$ units and $B/2$ amplification, followed by a void of $O-p$ units, followed by another share of $p$ units and $B/2$ amplification. This, however, only applies to its rewards from future blocks; for its reward from its own block, the calculation is as described above assuming it is entirely before the void.

\section{PPLNS variants}\label{sec:pplnsvar}
The PPLNS method described in \autoref{sec:pplns} is the standard way to enable hopping-proof pools without any operator risk. It has several variants that maintain this core nature.

One variant is to use a different decay function. Using an exponential function is equivalent to DGM with $o=1$. This gives a value of $\frac12(pB)^2$ for the product of the variance and maturity time, just like a step function. More generally, any decay function $r(x)$ can be used with the framework of \autoref{sec:framework} setting $O=0$. The invariant is then
\[\frac{\int_0^{\infty}r(x)x\ dx\cdot\int_0^{\infty}r(x)^2\ dx}{\left(\int_0^{\infty}r(x)\ dx\right)^3}(pB)^2.\]
The best result is obtained with a linear function $r(x)=[0\le x\le X]\cdot2(X-x)/X^2$, for which the product is $\frac49(pB)^2$.

In pay-once-PPLNS, every share is paid at most once. After a share is paid it is deleted, giving a chance to older shares to be paid for future blocks (a partially paid share will be partially deleted). A rapid succession of found blocks will pay an increasingly large part of the backlog of shares. If there are no remaining shares, block rewards stay with the operator (this is a probabilistic loan as with PPLNS, which is paid back in the form of cashing out pending scores if the pool ceases operating). $X$ needs to be less than 1; as $X\to1$, the variance becomes 0 and the maturity time and total loan become infinite.

Shift-PPLNS allows approximating standard PPLNS without having to store every share submitted, while keeping it hopping-proof. Shares are grouped into shifts. The pool keeps track of the total score for each participant in every shift, where every share submitted adds $pB$ to the score for the corresponding shift. The choice when to end one shift and start the next can be done freely, as long as it is completely independent of whether blocks are found. Generally a shift length $X$ will be chosen and a shift will end when it has accumulated $X$ units (every share has $p$ units). In addition a positive integer $N$ will be chosen representing the number of last shifts to pay. Whenever a shift ends, every participant is paid $\frac{SL}{NX_1}$, where $X_1$ is the total units in the shift that just ended, $L$ is the number of blocks found in that shift, and $S$ is the total score of the user in the previous $N$ shifts.

\chapter{Attack vectors}
\section{Pool-hopping}
A hoppable pool is one where the attractiveness of mining, in terms of expected earnings, variance and maturity time, varies according to the pool's current state. Pool-hopping is then the exploitation of this circumstance by mining only when the attractiveness is high and leaving when the attractiveness is low. This damages the overall attractiveness for participants who mine continuously.

Continuously mining in a hoppable pool is thus clearly disadvantageous; but a scenario where everyone optimally pool-hops is also not sustainable. Any hoppable pool will eventually reach a state when it is less attractive than hopping-proof pools; when this happens everyone will stop mining there, and it will forever remain in this state. The only sustainable system is one where everyone mines solo or at a pool which is hopping-proof, or at least very highly hopping-resistant.

The traditional form of hopping is to exploit the higher expected reward for mining in young rounds in pools which use the proportional method, and to a lesser extent, slush's method. This was discussed in \autoref{sec:prop} and \autoref{app:hop}.

Hopping can also be done in pools which base payments on some sort of buffer. Generally, the lower the buffer, the less attractive it is to participate. In SMPPS, for example, pool-hoppers can reduce their maturity time by mining only when the buffer is positive.

In some naive score-based methods which take temporal factors (rather than only shares and blocks submitted) into consideration in determining a share's score, it is possible to pool-hop based on fluctuations in the pool's hashrate. Generally, it is more profitable to mine when the pool's current hashrate is higher than average. This is analyzed in \autoref{app:fluctuate}.

\section{Block withholding}
The analysis so far rested on the assumption that the atomic action that can be carried out by participants in a pool is to perform the work needed to find a share, and submit it to the pool unconditionally. However, with the current protocol, miners can determine whether a share they have found is a valid block or not, and refrain from or delay submitting of blocks. This can be used for two kinds of attack, sabotage and lie in wait.
\subsection{Sabotage}
The simpler attack is sabotage, where the attacker never submits any blocks. This has no direct benefit for the attacker, only causing harm to the pool operator or participants.

Using a reward system without operator risk like PPLNS, each participant (including the attacker) will lose a portion of his rewards equal to the attacker's portion of the pool's hashrate. If the attacker's hashrate is $h$, the pool's total is $H$ and the pool fee is $f$, participants will obtain on average $(1-f)(1-h/H)pB$ per share. This can create a significant loss, but since it is difficult to detect, it will likely not cause desertion of the pool or any other long-term disruptions.

With a reward system that puts the operator at risk like PPS, the entire loss falls on the operator. He will gain on average $(f-h/H)pB$ per share submitted, and since $f$ is typically only a few percent, this can easily be negative. This way it is possible to cause bankruptcy of the pool.
\subsection{Lie in wait}
Lie in wait is a profitable attack where a miner postpones submitting blocks he found, and uses the knowledge of the imminent block to focus his mining on where it is most rewarding.

Let $T_0 = 10\,\textrm{min}$ be the average time to find a block in the network. Assume for example there are $m$ PPLNS pools with window parameter $X=1$ and hashrate of $H$ each, and that the attacker mines with a hashrate of $h$ in each (the waiting phase). Whenever he finds a block, he directs all his mining capacity to the pool where it is found (the ambush phase), and only after a period of time $T$ he submits the block and goes back to the waiting phase of mining in all $m$ pools. If another block is found anywhere before this time (making his own block invalid), he goes back to waiting.

The chance of a successful ambush (nobody else finding a block before time $T$) is $\exp(-T/T_0)\approx1-T/T_0$. During the ambush he will find on average $mhT2^{-32}$ shares, worth each roughly $pB$ above normal, because it will be rewarded for both the known imminent block and any future blocks. However, delaying the block reporting decays his existing shares in the pool by a total value of roughly $(pBhT/H)(H+(m-1)h)\approx pBhT2^{-32}$. If the ambush is failed, he will not gain any extra but will still have his shares decayed; the time for another block to be found is on average roughly $T/2$, so the lost value is $pBhT2^{-33}$. The expected gain from the ambush is therefore
\[pBh2^{-32}((1-T/T_0)(mT-T)-(T/T_0)T/2)\]
This is maximized when $T=\frac{m-1}{2m-1}T_0$ (some of the approximations used rely on the value of $T$ being small, so the actual optimal argument and value may be lower) and the maximum is $pBh2^{-32}\frac{(m-1)^2}{4m-2}T_0\approx pBh2^{-34}mT_0$. The total network hashrate is $H_0=2^{32}D/T_0$ so this is equal to $\frac{mhB}{4H_0}$. Each share found by the attacker has a probability of $p$ to be a valid block and enable an ambush, and thus the expected reward per share found, instead of $pB$, is $pB\left(1+\frac{mh}{4H_0}\right)$ (where it should be noted that $mh$ is the attacker's hashrate).

\subsection{Proposed solution -- Oblivious shares}
One workaround for block withholding attacks is the ``pop quiz'' -- occasionally, the pool will provide miners with work for which a solution is known, and flag participants who do not submit it. However, this leaves much to be desired with both precision and recall. A true solution is to modify the Bitcoin protocol to allow oblivious shares -- shares which, when found by miners, cannot be identified as a valid block with submitting to the pool for review. A possible way to do this will be as follows:
\begin{itemize}
\item{Every block will have 3 additional field associated with it -- SecretSeed, ExtraHash and SecretHash.}
\item{ExtraHash must be the hash of SecretSeed.}
\item{ExtraHash will be a part of the block header and will be one of the fields used in the calculation of the block hash.}
\item{SecretHash must be the hash of the concatenation of the block hash and SecretSeed.}
\item{For the block to be valid: Instead of requiring that the block hash is less than $2^{256}/(2^{32}D)$, it is required that the block hash is less than $2^{256} / 2^{32}$ and that SecretHash is less than $2^{256}/D$.}
\end{itemize}
The pool operator will choose SecretSeed and keep it secret. He will calculate ExtraHash and provide it to miners along with the other fields that go into the block hash. The miners can calculate the block hash and see if it is less than $2^{256} / 2^{32}$ (which happens with probability $2^{-32}$) and in this case submit it as a share. They do not know if this is a valid block because they don't know SecretSeed and cannot calculate SecretHash. The operator knows SecretSeed, calculates SecretHash, and if it is less than $2^{256}/D$ (with probability $p$) this is a valid block and it is broadcast to the network.

\chapter{Nonstandard reward systems}
\section{Shares as a future payment contract}
The reward methods discussed so far are coherent schemes for distributing the block rewards of the pooled mining among participants. All participants are treated equally, and unless the method is such that the operator absorbs variance by introducing a variable block fee, a fixed portion of the rewards are distributing among the miners one way or another.

A more general framework for considering the operation of mining pools is this: The pool operator buys shares from participants, and by default earns the right to enjoy the rewards if the share turns out to be a valid block. The participants, in return, receive either an immediate one-time payment, or a contract that entitles the participant to eventual payment based on some future contingency. Typically, to hedge his risks, the operator will offer contracts for which the payments depend on the future luck of the pool. The questions to ask then are how much each share is worth for the operator, how valuable is a specific contract for the participant, and what liability does the contract represent to the operator.

Every standard reward system is a special case of this, where the contracts simply indicate that the participant is entitled to payouts depending on future blocks found according to the method used. However, the more general setting allows more flexibility, as we will explore in the rest of this chapter.

\section{Variable block rewards}
The discussion so far has mostly assumed that the reward per block, $B$, is fixed. However, in reality the block reward is composed of two components, generation of new bitcoins and transaction fees. The bitcoin generation per block is cut in half every 210000 blocks, and the transaction fees vary rapidly based on the currently available transactions. As transaction fees become an ever larger portion of the block rewards, the variability of the rewards will become more significant.

In the standard reward method mindset, especially with methods which do not expose the operator to risk, it is tempting to think that the distributed rewards should be proportional to the value of each block found. After all, the premise is to distribute among participants the actual rewards obtained by the pool. However, this approach does not hold up when the reward is highly variable. The expected value of each share is equal to its probability to be a valid block multiplied by what the reward would be if it is, both evaluated at the time it is generated. This means that there will be pools able to offer share rewards based on the current potential block reward at the time of submission. However, if the reward given per share is contingent only on the value of future blocks found, then the expected reward is more or less independent of the current block reward. This means that pools of the latter kind are hoppable -- it is more attractive to mine when the current block reward is lower than average, and less attractive when it is higher.

To be hopping-proof, the pool must offer per share a contract with expected value proportional to the expected value of the share, which itself is proportional to the block reward at the time of submission. Most of the hopping-proof methods discussed were presented with this in mind -- the block reward is directly incorporated into the score rewarded for each share, so this only requires to make sure that the value used corresponds to what the block reward would be for this share if it turns out to be a valid block. In other methods the presentation deferred the incorporation of block rewards to the payment phase -- these merely require moving the multiplication by $B$ to the individual share level.

This applies to all reward systems, including PPS. To remain hopping-proof, PPS pools should not literally offer a fixed payout per share known in advance, but rather a payout of $(1-f)pB$ per share where $f$ is known in advance and $p$ and $B$ are the relevant values at the time each share is submitted.

\section{Hybrid reward methods}
We have seen that shares can be sold to the pool operator for a contract to be paid according to some specified reward system. For example, a contract to be paid strictly according to PPLNS poses little variance to the operator, but some variance and maturity time to the participant. A contract to be paid some deterministic amount immediately (as in PPS) puts the operator at some risk, and the utility cost of this risk will be reflected in the price offered per share. As an alternative, the pool could offer for every share, say, 70\% of the PPLNS reward and 30\% of the PPS reward. This allows creating new profiles for the probability distribution of rewards. The tradeoff that can be obtained in this way in terms of expectation, variance, maturity time and so on are comparable to what can be achieved using a standard parameterized method like DGM.

\section{Heterogeneous pools}
The treatment of share rewards as contracts leads us to the realization that the same contract does not have to be used for each participant. Each participant may be looking for a different tradeoff between the reward attributes, but he should not have to choose different pools for different reward profiles -- the same pool can offer different contracts based on the requirements of each participant.

The simplest implementation will be to offer payments based on a specific standard reward method, where each participant can customize his own parameters. The operator will have to scale the expected value of each contract based on its effect on the utility of his entire portfolio. In particular, more diverse contracts create higher operator variance, and the fee collected should take this into account.

Note that each share is rewarded as if all shares submitted to the pool used the same reward method and parameters as it. For example, if the contract specifies that the share's score decays by a factor of $r$ for every share submitted, it decays for \emph{every} share submitted, not just for shares that also use this contract. Likewise, it is rewarded for every block found, even if the block finder uses a different contract.

\section{Variable difficulty shares}
So far we have assumed that participants submit shares which match the hash target of difficulty 1. This choice is completely arbitrary -- for any value $d$, hash targets corresponding to difficulty $d$ can be used as shares. Using shares of higher difficulty reduce server load, but increase share-based variance.

Every method discussed can be trivially adapted to using a different share difficulty. The only requirement is that wherever $p=1/D$ is used, the value $p'=d/D$ is used instead, representing the probability that a share will be a valid block. Also, in some of the analysis the value $D'=D/d=1/p'$ takes the role of $D$.

It is also possible for different participants to use different share difficulties. Miners with a higher hashrate are less affected by share-based variance and so will be able to use more difficult shares. The pool could offer a discount on the fee for such shares in recognition of the reduced load. Every share should be scored based on its own difficulty, and so should its effect on share decay\footnote{For example, in the geometric method, each share submitted causes an update $s=s\left(1-p+\frac{p}{c}\right)$; the value of $p$ should be $d/D$ where $d$ is the value used for the share just submitted.}. This works especially well with the general framework of section \autoref{sec:framework}.

It is important that the share difficulty used for a participant will be decided before he is given work, otherwise he can collect rewards higher than the value of his work. For example, suppose a reward of $(d/D)B$ per share is offered, where the participant can choose after the fact whether to use $d=1$ or $d=2$. After computing $2^{32}$ hashes, he is expected to find one difficulty-1 share, and each such share has $1/2$ probability to also match the difficulty-2 target. By submitting each share according to the highest difficulty it matches, he will obtain on average $(1/2)(1/D)B+(1/2)(2/D)B=(3/2)(B/D)$ for this work, which is only worth $B/D$.\footnote{In general, if the participant can freely choose a posteriori a difficulty among $d_1, d_2, \ldots$, where the sequence is increasing and either its last element or its limit is $\infty$, the amplification factor is $\sum_{i}\left(1-\frac{d_i}{d_{i+1}}\right)$.}

\section{Proxy mining}
Normally, pools operate their own Bitcoin network node which is used to generate block headers crediting the pool with generation rewards. Alternatively, the pool can act as a proxy to another pool, requesting work and forwarding it to miners. Instead of receiving its income from block rewards, the frontend pool receives income from the shares it submits to the backend pool. This can also be seen as the frontend pool buying shares from miners and reselling them to the backend pool, perhaps using a different contract. The relation of this to normal pools is analogous to the relation of mining for a pool to solo mining.

One simple way this can work is if the backend pool is PPS, and the frontend pool also offers PPS under the same terms. This way there is no risk for the frontend operator, as everything he pays for shares he receives back from the backend pool. Such a pool also provides no benefits in terms of the reward profile, but it can provide benefits in marketing, interface features, decentralization and so on.

Somewhat more interesting is the case where the backend pool offers PPS at a high share difficulty, and the frontend offers PPS at a lower share difficulty (normal PPS pools are a special case where the backend pool is the Bitcoin network itself, offering PPS with a share difficulty of $D$). It could be that the backend pool caters to larger miners and has too much variance for participants with low hashrate. Those could be serviced by the frontend pool. If the share difficulty of the backend pool is $d$, then the frontend pool's variance is $(d/D)$ times the variance of a normal PPS pool, so it can safely operate with only a small markup over the backend pool's fee.

A more interesting scenario is when the frontend pool offers PPS but the backend pool does not. This way, the backend pool will typically require a very low fee, and with the variance reduction it enables, the frontend can safely offer PPS, with all its advantages, at a low fee. If the total hashrate of the backend pool is $H$ and of the frontend pool is $h$, then the frontend operator's variance is $(h/H)$ times the variance of a normal PPS pool.

One promising candidate to act as a backend pool is \emph{p2pool}, a distributed mining pool (\cite{p2pool}). It can grow to a large size (enabling high variance reduction) without risking concentration of power or distributed denial of service (DDoS) attacks.

The frontend pool can be a proxy to more than one backend pool, and this can help further reduce the variance. If there are $m$ pools (which do not absorb any variance themselves) with respective hashrates $H_1, H_2, \ldots$ and total hashrate $H=\sum_iH_i$, the frontend pool should request work from all pools and distribute its hashrate among the pools in proportion to each pool's hashrate. This way the variance reduction factor is $(h/H)$, better than what can be achieved with any single pool. This also helps resist DDoS attacks against the backend pool.

The frontend pool can offer payment systems other than PPS. However, to reduce risk, the contracts will have to be conditioned on the pool's income, which depends on the blocks found by the backend pool. This adds complexity which makes this arrangement unattractive, but it could still be useful to adapt the backend pool to a different reward system or usability features.

A miner can act as his own proxy pool, distributing his hashrate proportionally among several pools to reduce variance. This only helps with pool-based variance, so if his hashrate is low, he will want to avoid pools with high share difficulty.

\section{Score cashout}
Every share submitted by a participant gives him a contract entitling him to future payments. In purely-score based reward methods like DGM, the total of all of a participant's contracts can be encoded with a single value, his score. Sometimes he will want to cash out his contract for some immediate payment. For example, if he stops mining and wants to make a clean break from the pool without having to wait for his payments to mature.

To this end, it will be useful if pools offer just such a feature -- cashing out score for immediate payment. Because of the risk this creates, the payment will typically be lower than the full expected reward for the score. This is reasonable because it is to be used in specific circumstances rather than as a matter of course. Also, as long as it is only used occasionally, the operator risk for offering it is minimal (and if it is used too much, the cashout fee can be increased).

\section{Score markets}
Allowing pool participants to sell contracts back to the operator leads us to the idea of selling contracts to other people, even if they are not miners themselves. Different people have different valuations for contracts; people who have an urgent need for money will have less value for high maturity time contracts, and people with large financial reserves will have more value than others for high variance contracts. Efficiency can be increased by trading contracts to where they are most useful. The pool operator is not affected by this; he will still have to offer the same payments to the owner of the contract, whoever that may be.

An exchange market can be created where traders can bid and ask for contracts with specified attributes. This can create some technical challenges as the value of a contract is rapidly changing. Alternatively, a trader could offer to buy contracts of a pool at a specified price agreed upon with the pool operator. If this is seamlessly integrated into the interface, participants can be offered PPS payments without knowing there is a 3rd party buying contracts, and without any risk for the operator. Because the one taking the risk is the buyer, who can have large financial reserves and only needs a rudimentary understanding of how mining works, this can allow the effective PPS fee to be low.

\section{Streamlined PPS investments}
Recall from \autoref{app:net} that the fee required for a normal PPS pool to operate safely is inversely proportional to its reserve. It follows that the pool will want as high a reserve as possible to be more attractive. Raising the capital needed for this can be challenging, but there is a way to streamline investing by any interested party.

The pool should allow anyone to create an investor account and deposit bitcoins in the pool. These funds will be used as the reserve. Whenever payment needs to be sent for a submitted share, it is deducted from each investor account in proportion to its balance. When a block reward is received, it is distributed among accounts in proportion to their balance. This way investors share the risk and reward, and can withdraw as necessary.

Depositing can be streamlined even further by allowing, instead of creating an account, to simply send bitcoins specifying a Bitcoin return address and a time to cash out. An internal account will be created with this balance, and whatever its balance is at the specified time is sent to the return address. The investor can then reinvest if he wishes.

The pool operator can create a virtual balance for himself as compensation for his work and costs, participating in the risks and rewards without investing his own capital. He can choose whichever value he wants, but this needs to be transparent -- if it is too high, investors may want to look for opportunities with a higher return on investment. Having a balance too high also puts excessive risk on the operator. The virtual balance should not be included in calculations that relate to the total reserve.

\chapter{Conclusion}
In this paper we have introduced Bitcoin mining and explained why the high variance in the rewards for this activity creates the need for mining pools. We have explored several methods used by pools to distribute rewards, and seen how some of them fall short of the ideal of fair reward distribution -- most notably due to the exploit known as ``pool-hopping''. We have developed a general framework for hopping-proof methods which are immune to this problem, and analyzed some specific examples. Finally, we have discussed some innovative ways pools could improve the service they provide going forward.

\appendix

\chapter{Properties of proportional pools with constant hashrate} \label{app:prop}
Let us consider a pool using the proportional system, with a fixed roster of miners with constant hashrate. For a share submitted at a random time, the number of shares in the round in which it is included, $N$, follows a (shifted) negative binomial distribution with stopping time 2 and failure rate $p$:
\[\mathrm{Pr}(N)=Np^2(1-p)^{N-1}\]
The reward for this share is $w=\frac{B}{N}$ (ignoring fees, which could be simply factored into $B$), so the expected payout is:
\[\mathbb{E}[w]=\sum_{N=1}^{\infty}\mathrm{Pr}(N)\frac{B}{N} = \sum_{N=1}^{\infty}p^2(1-p)^{N-1}B=p^2B \sum_{K=0}^{\infty}(1-p)^K=pB\]
The expected squared payout is:
\[\mathbb{E}[w^2]=\sum_{N=1}^{\infty}\mathrm{Pr}(N)\left(\frac{B}{N}\right)^2 = \frac{p^2B^2}{1-p}\sum_{N=1}^{\infty}\frac{(1-p)^N}{N} = \frac{-p^2B^2\ln p}{1-p}\]
Where the last equality follows from the power series expansion of the natural logarithm around 1. So the variance is:
\[\mathbb{V}[w]=\mathbb{E}[w^2]-\mathbb{E}[w]^2 = \frac{-p^2B^2\ln p}{1-p} - (pB)^2 = -p^2B^2\left(1+\frac{\ln p}{1-p}\right)\]
This is roughly $p^2B^2(\ln D-1)$ or even more crudely $p^2B^2\ln D$. The variance per share of solo mining is $p(1-p)B^2$ or roughly $p B^2$, so this represents an improvement by a factor of $\frac{D}{\ln D}$. For example, if the difficulty is $D=1.5\cdot10^6$, the variance per share in the pool is $1.13\cdot10^5$ times less than solo.

\chapter{Pool-hopping in proportional pools} \label{app:hop}
The expected payout for a newly submitted share depends on the number of shares already submitted in the round. The total number of shares in a round, $N$, follows the geometric distribution with success parameter $p$:
\[\mathrm{Pr}(N)=p(1-p)^{N-1}\]
Given that $I$ shares have already been submitted in the current round, we know that $N>I$. Conditioning on this, we have
\[\mathrm{Pr}(N|N>I)=\begin{cases}0&N\le I\\p(1-p)^{N-I-1}&N>I\end{cases}\]
The reward of a share is $w=\frac{B}{N}$, so the expected payout for a share submitted at this time is:
\[\mathbb{E}[w|N>I]=\sum_{N=I+1}^{\infty}\frac{p(1-p)^{N-I-1}B}{N}\]
We can evaluate this expression for any given value of $I$. In particular it is interesting to note that for $I$ which is only a few shares, this is roughly $pB\ln D$; so shares submitted at the very start are worth $\ln D$ times the normal reward. A closed-form expression for a general $I$ requires the use of relatively obscure nonelementary functions such as the Hurwitz Lerch transcendent. We can obtain a more friendly form by estimating this as an integral, using the exponential integral $\mathrm{E}_1$:
\[\mathrm{E}_1(x)=\int_{1}^{\infty}\frac{\exp(-xt)}{t}\ dt\]
Denoting $x=pI,\ y=pN$ -- the already submitted shares and the total number of shares, respectively, expressed as a fraction of the difficulty -- we have:
\[\mathbb{E}[w|y>x]\approx\int_{x}^{\infty}\frac{p(1-p)^{(y-x)/p}B}{y}\ dy\approx \exp(x)\mathrm{E}_1(x)pB\]
So, the function $f(x)=\exp(x)\mathrm{E}_1(x)$ represents the amplification factor when $xD$ shares have already been submitted. This is a monotonically decreasing function, and we can gain some insight into the behavior of this function by looking at its asymptotics -- for $x\approx0$ we have $f(x)\approx-\ln x-\gamma$, where $\gamma$ is the Euler gamma constant; and for $x\approx\infty$ we have $f(x)\approx\frac{1}{x+1}$.

It is also interesting to note when do we have $f(x)=1$. This happens for $x_0 = 0.4348182...$, or approximately 43.5\%.

When a pool-hopper has the choice between mining in a given proportional pool or mining solo (or in a fair score-based pool), he should mine for the pool if $x<x_0$ (with amplification $f(x)$), and solo otherwise (with amplification 1). The value of $x$ at a random time follows the exponential distribution with mean 1 (assuming the pool has constant hashrate, which relies on the effect of the hopper on the pool being negligible), so the expected amplification at a random time is:
\[\mathbb{E}\left[\frac{w}{pB}\right]\approx \int_0^{x_0}\exp(-x)f(x)\ dx+\int_{x_0}^{\infty}\exp(-x)\ dx \approx 1.28149...\]
So, following this strategy, the pool-hopper can obtain 128\% of the normal payout for his given hashrate. Of course, this assumes perfect knowledge of the age of the round at any time, and flawless execution of the strategy; but the figure in practice should be very close.

What if there are more than one proportional pool? The hopper will do well to always mine for the pool with the youngest round, unless all pools have more than $x_0D$ shares, in which case he should fall back to solo. The minimum value of $x$ among $m$ pools, assuming they all have constant hashrate (though not necessarily equal between them), follows the exponential distribution with mean $1/m$. So the expected amplification is:
\[\mathbb{E}\left[\frac{w}{pB}\right]\approx m\left(\int_0^{x_0}\exp(-mx)f(x)\ dx+\int_{x_0}^{\infty}\exp(-mx)\ dx\right)\]
The values of this for various values of $m$ are listed in \autoref{tab:hop}.
\\
\begin{table}
\caption{Amplification factor of pool-hopping, as a function of the number of proportional pools.}
\begin{tabular}{c||c|c}
$m$&with fallback&without fallback\\
\hline
1& 1.28149& 1\\
2& 1.5159& 1.38629\\
3& 1.71404& 1.64792\\
4& 1.88393& 1.84839\\
5& 2.03152& 2.0118\\
6& 2.16131& 2.15011\\
7& 2.27669& 2.27023\\
8& 2.38028& 2.3765\\
9& 2.4741& 2.47188\\
10& 2.55975& 2.55843\\
\hline
15& 2.90159& 2.90148\\
20& 3.15341& 3.1534\\
25& 3.353& 3.353
\end{tabular}
\label{tab:hop}
\end{table}

When $m$ is large, the need for fallback mining diminishes; it is possible to simply always mine for the pool with the youngest round, even if it is older than $x_0$. It is also simpler to evaluate the amplification of this strategy. We have:
\[\mathbb{E}\left[\frac{w}{pB}\right]\approx m\int_0^{\infty}\exp(-mx)f(x)\ dx = \frac{m\ln m}{m-1}\]
These values also appear in \autoref{tab:hop}.

How does this affect the honest miners in a proportional pool? In the worst case, there are infinitely many pool-hoppers, so whenever a block is found by the pool, the hoppers will instantly bring it to a state of $x_0D$ shares. For a share submitted at a random time, the eventual total number of shares submitted by continuous miners in the block follows the same distribution as in the hopper-free scenario; however, in addition to these shares there will always be $x_0D$ shares submitted by hoppers. Hence, the expected payout per share, using the continuous approximation, is
\[\mathbb{E}\left[\frac{w}{pB}\right]\approx\int_0^{\infty}\frac{x\exp(-x)}{x+x_0}\ dx = 1-x_0f(x_0) = 1-x_0 = 0.56518...\]
So their expected payout will only be 56.5\% of the fair reward.

\chapter{Safety nets for PPS pools} \label{app:net}
In this appendix we discuss the financial reserves that should be kept by the operator of a PPS pool in order to assure its proper functioning. The reserve should be large enough to make sure that the probability of the pool ever going bankrupt is low.

Assuming there are no sabotage attacks, invalid blocks and so on, the operator's balance can be modeled as the Markov chain
\[X_{t+1}-X_t = \left\{\begin{array}{ll}-(1-f)pB+B&\mathrm{w.p.}\ p\\-(1-f)pB&\mathrm{w.p.}\ 1-p\end{array}\right.\]
where each share submitted corresponds to a step. Each difference has expectation $fpB$ and variance approximately $pB^2$, so by the central limit theorem, the long-term behavior of this stochastic process is equivalent to that of
\[X_{t+1}-X_t = \left\{\begin{array}{ll}+\sqrt{p}B&\mathrm{w.p.}\ (1+f\sqrt{p})/2\\-\sqrt{p}B&\mathrm{w.p.}\ (1-f\sqrt{p})/2\end{array}\right.\]
which has the same expectation and variance. This, in turn, is equivalent to
\[X_{t+1}-X_t = \left\{\begin{array}{ll}+1&\mathrm{w.p.}\ (1+f\sqrt{p})/2\\-1&\mathrm{w.p.}\ (1-f\sqrt{p})/2\end{array}\right.\]
where the initial condition has been scaled by a factor of $\sqrt{p}B$.

We will let $a_n$ denote the probability that we will ever reach $0$ (representing bankruptcy of the pool) given that we start in state $n$. By conditioning on the first step we get, denoting $q=(1+f\sqrt{p})/2$,
\[a_n=qa_{n+1}+(1-q)a_{n-1}.\]
The characteristic polynomial of this recurrence equation is $q\lambda^2-\lambda+(1-q)$, and thus it has the general solution $a_n = A+B((1-q)/q)^n$. Substituting the boundary conditions $a_0=1,\ a_{\infty}=0$, we have $A=0,\ B=1$ and so
\[a_n=\left(\frac{1-q}{q}\right)^n=\left(\frac{1-f\sqrt{p}}{1+f\sqrt{p}}\right)^n\approx\exp(-2fn\sqrt{p}).\]

If the operator starts with a reserve of $R$, the probability that the pool will ever go bankrupt is
\[\delta = a_{R/(\sqrt{p}B)} \approx \exp\left(\frac{-2fR\sqrt{p}}{\sqrt{p}B}\right) = \exp\left(\frac{-2fR}{B}\right).\]
Conversely, to maintain a bankruptcy probability of at most $\delta$, the pool should keep a reserve of at least
\[R=\frac{B\ln\tfrac1\delta}{2f}.\]

For example, if $B=50\,\textrm{BTC},\ \delta=1/1000$ and $f=0.05$ (5\% fee), the reserve should be
\[R=\frac{50\,\textrm{BTC}\cdot\ln1000}{2\cdot0.05}\approx 3454\,\textrm{BTC}.\]
If the operator tries to make do with $f=0.01$ but only has a reserve of $500\,\textrm{BTC}$, then
\[\delta=\exp\left(\frac{-2\cdot0.01\cdot500\,\textrm{BTC}}{50\,\textrm{BTC}}\right)\approx0.819\]
so he has 81.9\% chance of eventual bankruptcy.

\chapter{The hopping immunity theorem}\label{app:immune}
\begin{thmHop}Suppose that the difficulty $D$ and block reward $B$ are fixed. Let a reward method distribute a fixed amount of $(1-f)B$ among shares in the round, according to a deterministic function of the round length and the share index. If the expected reward per share at the time it is submitted is always $(1-f)pB$, then the entire reward is always given to the last share submitted.\end{thmHop}
\begin{proof}By assumption, the reward for the $i$th share in a round of length $N$ is given by a function $f(i,N)$ with the following properties:
\begin{enumerate}
\item{$f(i,N)\ge0$ for every $i,N\in\mathbb{N},\ i\le N$.}
\item{$\sum_{i=1}^Nf(i,N)=(1-f)B$.}
\item{If $N$ follows the geometric distribution with success probability $p$ then
\[\mathbb{E}[f(I,N)|N\ge I]=(1-f)pB.\]}
\end{enumerate}
We will show by complete induction on $I$ that $f(I,N) = \delta_{IN}(1-f)B$.\footnote{$\delta_{ij}$ is the Kronecker delta, $\delta_{ij} = \left\{\begin{matrix}
1, & \mbox{if } i=j   \\
0, & \mbox{if } i \ne j   \end{matrix}\right.$} By the fixed reward assumption, $\sum_{i=1}^If(i,I)=(1-f)B$, and by the induction hypothesis, $f(i,I)=0$ for all $i<I$. Therefore, $f(I,I)=(1-f)B$. By the expected value assumption we have
\[(1-f)pB=\sum_{N=I}^{\infty}p(1-p)^{N-I}f(I,N)=pf(I,I)+\sum_{N=I+1}^{\infty}p(1-p)^{N-I}f(I,N)\]
\[\sum_{N=I+1}^{\infty}p(1-p)^{N-I}f(I,N)=(1-f)pB-pf(I,I)=0\]
By the nonnegativity assumption, $f(I,N)\ge0$ and therefore $f(I,N)=0$ for all $N>I$. \end{proof}

\chapter{Properties of the geometric method} \label{app:geom}
In this appendix we will analyze the expectation, variance and maturity time\footnote{``Maturity time'' is used here to mean the average number of shares, in multiples of the difficulty, until a given quantity of reward is confirmed.} of the reward for a share submitted to a pool using the geometric method. These are all independent of any history of the pool, and the expected reward is independent of any future changes in the difficulty or the block reward, making the method hopping-proof.

We will first analyze the case that the difficulty $D$ (and its inverse $p$) and block reward $B$ are fixed. If $I$ shares have already been submitted in the current round, the distribution of the eventual total number of shares in the round is
\[\mathrm{Pr}(N|N>I)=\begin{cases}0&N\le I\\p(1-p)^{N-I-1}&N>I\end{cases}\]
The value of $s$ at this point is $r^I$, so the addition to the score for a submitted share is $r^IpB$. The value of $s$ at the end of the round is $r^N$, so the payout for this share is $w=\frac{(1-f)(r-1)S}{sp}=(1-f)(r-1)r^{I-N}B$. The expected payout is
\[\mathbb{E}[w]=\sum_{N=I+1}^{\infty}p(1-p)^{N-I-1}(1-f)(r-1)r^{I-N}B = (1-f)pB\frac{r-1}{r}\sum_{N=0}^{\infty}\left(\frac{1-p}{r}\right)^{N}\]
We evaluate the sum using $r=1-p+\frac{p}{c}$ from which follows $1-p=\frac{1-cr}{1-c}$:
\[\sum_{N=0}^{\infty}\left(\frac{1-p}{r}\right)^{N}=\frac{1}{1-\frac{1-p}{r}}=\frac{r}{r-(1-p)}=\frac{r}{r-\frac
{1-cr}{1-c}}=\frac{r(1-c)}{r(1-c)-(1-cr)}=\frac{r(1-c)}{r-1}\]
So
\[\mathbb{E}[w]=(1-f)pB\frac{r-1}{r}\cdot\frac{r(1-c)}{r-1}=(1-f)(1-c)pB.\]
The expression for the variance of the payout is too cumbersome to be derived here. However, with the aid of a computer algebra system, we find that
\[\mathbb{E}[w^2]=\sum_{N=I+1}^{\infty}p(1-p)^{N-I-1}((1-f)(r-1)r^{I-N}B)^2=\frac{((1-f)(1-c)pB)^2}{p+c(2-c)(1-p)}\]
\[\mathbb{V}[w]=\mathbb{E}[w^2]-\mathbb{E}[w]^2=\frac{(1-c)^4(1-f)^2(1-p)p^2B^2}{p+c(2-c)(1-p)}\approx\frac{(1-c)^4(1-f)^2p^2B^2}{p+c(2-c)}\]
Assuming that $f\approx0,\ c\approx0$, we have $\mathbb{V}[w]\approx\frac{(pB)^2}{p+2c}$, which is a reduction by a factor of $1+\frac{2c}{p}$ over the solo variance of $pB^2$ per share. If $c\not\approx0$ but the average fee $c+f-cf\approx0$ then $\mathbb{V}[w]\approx\frac{(1-c)^2(pB)^2}{c(2-c)}$.

As with other reward systems, the rewards for proximate shares are correlated, so the total variance for submitting several shares scales worse than linearly. However, in pools that do not involve operator risk, The minimal total variance that can be achieved for a participant who constitutes a proportion $q$ of the pool is $q$ times the solo variance. In contrast, the geometric method with high $c$ allows achieving less variance than this, and in the limit $c\to1$ the variance is 0 per share regardless of the participant's hashrate.

The total reward given to all participants sums to a value which gives the operator an effective score of $\frac{1}{r-1}$. The variable fee $w_0$ is therefore equal to $\frac{1}{r-1}$ times the reward for the first share submitted, and we can leverage this to calculate the statistical properties of this fee. The expected variable fee per block is
\[\mathbb{E}[w_0]=\frac{(1-f)(1-c)pB}{r-1}=\frac{(1-f)(1-c)pB}{p(1-c)/c}=c(1-f)B\]
Adding to this a fixed fee of $fB$, the total fee is on average $(c+f-cf)B$ per block.
The variance is
\[\mathbb{V}[w_0]=\frac{(1-c)^4(1-f)^2(1-p)p^2B^2}{p+c(2-c)(1-p)}\left(\frac{c}{p(1-c)}\right)^2=
\frac{(1-c)^2(1-f)^2(1-p)c^2B^2}{p+c(2-c)(1-p)}\approx\frac{c}{2-c}B^2\]

The \emph{maturity time} is defined as
\[T = \frac{p\sum_{N=0}^{\infty}N\mathbb{E}[w;N]}{\sum_{N=0}^{\infty}\mathbb{E}[w;N]}\]
Where $w;N$ is the additional reward due for the current share after $N$ more shares were submitted (the denominator is simply the expected total reward). In the geometric method reward is due if the future share is the one that ends the round, which happens with probability $p(1-p)^N$, and in this case the reward is $(1-f)(r-1)r^{-N-1}B$, so $\mathbb{E}[w;N] = (1-f)(r-1)(1-p)^Nr^{-N-1}pB$ and
\[T = \frac{p\sum_{N=0}^{\infty}N\mathbb{E}[w;N]}{\sum_{N=0}^{\infty}\mathbb{E}[w;N]} = c(1-p) \approx c\]
Thus, while decreasing $c$ significantly increases the variance, it has the advantage that if a reward is received at all, it is received quickly.

To show that the expected payout per share is always $(1-f)(1-c)pB$ even in the face of future changes in difficulty and block reward, we will prove the following invariant: A participant with a current score of $S$ is expected to receive a reward of $(1-f)(1-c)\frac{S}{s}$ (which we will call the \emph{effective score}) for it. Since the score given for a share is $spB$, it will follow that the expected reward per share is $(1-f)(1-c)pB$.

Clearly the effective score tends to 0 as shares are submitted, and the only time when either the effective score or the reward change is when a share is processed, so it is enough to show that the expected change in the effective score is equal to the negative of the expected change in the reward.

Suppose a share is submitted when the current value of $s$ is $s_0$. With probability $p$ (which need not be equal to the value of $p$ that was current when the participant in question accumulated his score) it will be a valid block. Before payment is given $s$ is updated to $s_1=s_0r$. The addition to the reward is $\frac{(1-f)(r-1)S}{ps_1}=\frac{(1-f)(r-1)S}{ps_0r}$, and the effective score is reduced to 0 which is a change of $-(1-f)(1-c)\frac{S}{s_0}$. With probability $1-p$ the share is not a valid block, meaning that the reward does not change but the effective score changes by
\[(1-f)(1-c)\frac{S}{s_1}-(1-f)(1-c)\frac{S}{s_0}=\frac{(1-f)(1-c)(1-r)S}{rs_0}.\]
The expected change in the reward is $\frac{(1-f)(r-1)S}{s_0r}$. The expected change in the effective score is
\begin{eqnarray*}
&&-p(1-f)(1-c)\frac{S}{s_0}+\frac{(1-p)(1-f)(1-c)(1-r)S}{s_0r}=\\
&=&\frac{(1-f)(1-c)S((1-r)(1-p)-pr)}{s_0r}=\frac{(1-f)S}{s_0r}(1-c)(1-p-r)=\\
&=&\frac{(1-f)S}{s_0r}(1-c)\frac{-p}{c}=-\frac{(1-f)(r-1)S}{s_0r}.
\end{eqnarray*}
This concludes the proof.

\chapter{Properties of *MPPS pools} \label{app:MPPS}

\section{Expected loss in MPPS}
Suppose Alice joins an MPPS pool, and her portion of the pool hashrate is $q$. After a total of $nD$ shares have been submitted to the pool (the amount it takes to find $n$ blocks on average), the amount submitted by Alice is $qnD$ so her PPS balance is $(qnD)\cdot(pB)=qnB$. The amount of blocks found by the pool during this time, $L$, follows the Poisson distribution with mean $n$, and Alice's proportional balance is $qLB$. The payment received is therefore $\min(qnB,qLB) = \min(n,L)qB$ where the fair reward is $nqB$. We have
\[\mathbb{E}[\min(n,L)] = \sum_{L=0}^{\infty}\frac{e^{-n}n^L}{L!}\min(n,L) =
\sum_{L=0}^{\lfloor n\rfloor}\frac{e^{-n}n^L}{L!}L + \sum_{L=\lfloor n\rfloor+1}^{\infty}\frac{e^{-n}n^L}{L!}n\]
which for integer $n$ is equal to $n\left(1-\frac{e^{-n}n^n}{n!}\right)\approx n\left(1-\frac{1}{\sqrt{2\pi n}}\right)$. So, the average portion of the reward lost to this method is $\frac{1}{\sqrt{2\pi n}}$.
\section{Maturity time in SMPPS}
We consider a participant who submits a share with due reward $w$, and track his received reward over time. We assume for simplicity that the buffer stays at a constant level throughout this time (this will happen if every round is of average length). When a block is found, he receives a payment of $\frac{wB}{-R}$ so his new due reward is $w\left(1-\frac{B}{-R}\right)$. This means that the due reward after $n$ blocks were found is $w\left(1-\frac{B}{-R}\right)^n$ so the payment given at that time is $w\left(1-\frac{B}{-R}\right)^{n-1} - w\left(1-\frac{B}{-R}\right)^n$. The maturity time is
\[\frac1w\sum_{n=1}^{\infty}n\left(w\left(1-\frac{B}{-R}\right)^{n-1} - w\left(1-\frac{B}{-R}\right)^n\right)=
\sum_{n=0}^{\infty}\left(1-\frac{B}{-R}\right)^n=\frac{-R}{B}.\]

\chapter{Hashrate fluctuation pool-hopping} \label{app:fluctuate}
In this appendix we will analyze pool-hopping in temporal score-based methods utilizing fluctuations in the pool's hashrate. As an example consider slush's method where each share submitted at time $T$ is given a score of $\exp(T/C)$, and let $H(T)$ be the pool's hashrate at time $T$. We will assume for simplicity that the current round is old enough that we can ignore its beginning and consider it to be at $T=-\infty$. Over a period of length $dT$ around time $T$, on average $\frac{H(T)dT}{2^{32}}$ shares are found; we will assume shares are continuous and that this is the exact number of shares found. Block finding is a non-time-homogeneous Poisson process with rate $\lambda(T)=\frac{H(T)}{2^{32}D}$; the probability that the first block found after time $0$ is at time at most $T$ is $1-\exp\left(-\int_0^T\lambda(t)\ dt\right)$. The probability density function of the first block found after time $0$ is therefore $P(T) = \lambda(T)\exp\left(-\int_0^T\lambda(t)\ dt\right)$. Given that this is equal to $T$, the payout for a share submitted at time $0$ is $\frac{B}{D\int_{-\infty}^T\lambda(t)\exp(t/C)\ dt}$. The expected payout is therefore
\[\int_0^{\infty}\frac{B\lambda(T)\exp\left(-\int_0^T\lambda(t)\ dt\right)}{D\int_{-\infty}^T\lambda(t)\exp(t/C)\ dt}\ dT.\]
The quantity $\bar{H}=\frac{\int_{-\infty}^0H(T)\exp(T/C)\ dT}{\int_{-\infty}^0\exp(T/C)\ dT}=\frac1C\int_{-\infty}^0H(T)\exp(T/C)\ dT$ is the effective average hashrate over the retrospective window considered by the method at time $0$. We will also denote $\bar{\lambda}=\frac{\bar{H}}{2^{32}D}$.

The future hashrate for $T>0$ is unknown, but can be estimated to be about the current hashrate $H(0)$, especially if $C$ is low enough that only a short period of time into the future needs to be considered. Then the expected payout for a share submitted at time 0 can be expressed with a hypergeometric function:
\[\int_0^{\infty}\frac{B\lambda(0)\exp\left(-T\lambda(0)\right)}{DC\bar{\lambda}+D\int_0^T\lambda(0)\exp(t/C)\ dt}\ dT=\frac{B\,{}_2F_1(1,1+\lambda(0);2+\lambda(0);1-\frac{\bar{\lambda}}{\lambda(0)})}{D(1+C\lambda(0))}.\]
This function increases with $\lambda(0)$ and decreases with $\bar{\lambda}$. For $\lambda(0)\approx\infty$, the method reduces to proportional with a round age of $C\bar{\lambda}$; the payout is $\exp(C\bar{\lambda})\mathrm{E}_1(C\bar{\lambda})$ times the normal reward. For $\lambda(0)\approx0$, the amplification is roughly $\frac{\lambda(0)}{\bar{\lambda}}\ln\left(\frac{\bar{\lambda}}{\lambda(0)}\right)$.

\newpage
\bibliographystyle{plain} 
\addcontentsline{toc}{chapter}{Bibliography}
\bibliography{pool_analysis}
\end{document}